\documentclass[journal,twoside,web]{ieeecolor}
\usepackage{tmi}
\usepackage{cite}
\usepackage{amsmath,amssymb,amsfonts}
\usepackage{textcomp}
\usepackage{mathtools}
\usepackage{hyperref}
\usepackage{algorithm}
\usepackage{algorithmic}
\usepackage{subcaption,graphicx}
\usepackage{bm}
\usepackage{soul}

\def\BibTeX{{\rm B\kern-.05em{\sc i\kern-.025em b}\kern-.08em
    T\kern-.1667em\lower.7ex\hbox{E}\kern-.125emX}}
\begin{document}
\title{Diffusion Models for Counterfactual Generation and Anomaly Detection in Brain Images}
\author{Alessandro Fontanella, Grant Mair, Joanna Wardlaw, Emanuele Trucco, Amos Storkey
\thanks{Alessandro Fontanella and Amos Storkey are associated with the School of Informatics, University of Edinburgh, Edinburgh, UK (e-mails: A.Fontanella@sms.ed.ac.uk; A.Storkey@ed.ac.uk). }
\thanks{Grant Mair and Joanna Wardlaw are associated with the Centre for Clinical Brain Sciences, University of Edinburgh, Edinburgh, UK, and with the Department of Clinical Neurosciences, NHS Lothian. 
Joanna Wardlaw is also associated with the UK Dementia Research Institute Centre, University of Edinburgh,  Edinburgh, UK (e-mails: grant.mair@ed.ac.uk; Joanna.Wardlaw@ed.ac.uk).}
\thanks{Emanuele Trucco is associated with the VAMPIRE project / CVIP, Computing, School of Science and Engineering, University of Dundee, UK (e-mail: E.Trucco@dundee.ac.uk).}}

\maketitle

\begin{abstract}
Segmentation masks of pathological areas are useful in many medical applications, such as brain tumour and stroke management. Moreover, healthy counterfactuals of diseased images can be used to enhance radiologists' training files and to improve the interpretability of segmentation models. In this work, we present a weakly supervised method to generate a healthy version of a diseased image and then use it to obtain a pixel-wise anomaly map. To do so, we start by considering a saliency map that approximately covers the pathological areas, obtained with ACAT. Then, we propose a technique that allows to perform targeted modifications to these regions, while preserving the rest of the image. In particular, we employ a diffusion model trained on healthy samples and combine Denoising Diffusion Probabilistic Model (DDPM) and Denoising Diffusion Implicit Model (DDIM) at each step of the sampling process. DDPM is used to modify the areas affected by a lesion within the saliency map, while DDIM guarantees reconstruction of the normal anatomy outside of it. The two parts are also fused at each timestep, to guarantee the generation of a sample with a coherent appearance and a seamless transition between edited and unedited parts. We verify that when our method is applied to healthy samples, the input images are reconstructed without significant modifications. We compare our approach with alternative weakly supervised methods on the task of brain lesion segmentation, achieving the highest mean Dice and IoU scores among the models considered.

\end{abstract}


\section{Introduction}
\label{sec:introduction}
The remarkable progress in advanced imaging technologies has led to a significant enhancement in the quality of medical care for patients. These cutting-edge tools empower radiologists to achieve ever-increasing levels of accuracy when diagnosing suspicious regions such as tumors, polyps, and areas of blood rupture~\cite{acharya1995}. Moreover, physicians are now able to implement precise and carefully measured treatment methods, thanks to the invaluable support provided by these imaging technologies. Indeed, the detection of pathological markers in medical images plays an important role in diagnosing disease and monitoring its progression. However, in many cases, segmentation of the Regions of Interest (ROI) is performed manually by radiologists, making it not only an expensive process but also prone to errors and inconsistencies across different annotators ~\cite{grunberg2017annotating, fontanellaclassification}. Therefore, the development of automated ROI detection systems is a very active area of research, for its potential to save time and money, while mitigating some of the inherent biases associated with human evaluations.

When a patient is diagnosed with brain tumour, segmentation of the pathological regions is important for planning the surgical treatments, monitoring the growth of the tumour and for image-guided intervention~\cite{ranjbarzadeh2021brain}. In particular, Magnetic Resonance Imaging (MRI) is a widely used non-invasive technique that generates a vast array of tissue contrasts. Medical experts have extensively employed it to diagnose brain tumors. However, the normal anatomy can be severely distorted by the tumour, making it harder to plan surgical approaches that avoid key structures. For this reason, generating an equivalent healthy image could improve surgical planning by helping the identification of anatomical areas.

\begin{figure*}[t]
\centering
\includegraphics[width=1\textwidth]{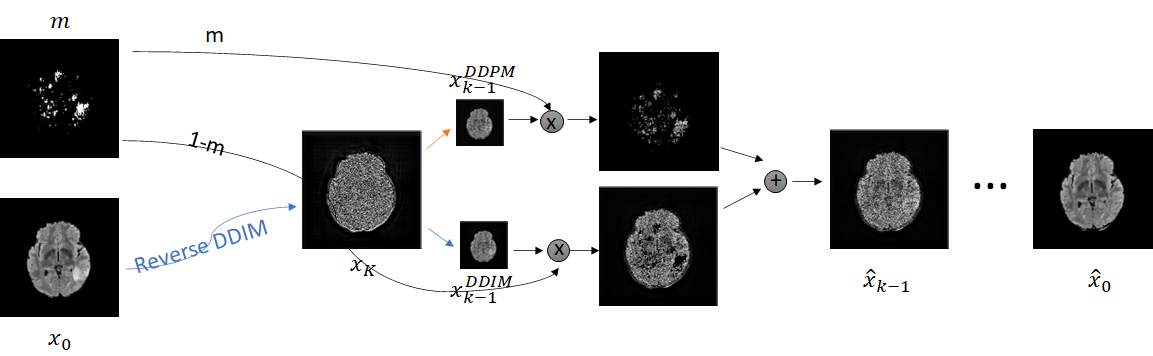}
\caption{Our approach begins by transforming an abnormal image $\bm{x}_0$ into its noised version $\bm{x}_K$ using the reversed sampling technique of DDIMs. Subsequently, we employ DDPM sampling to modify the pathological area, identified through the saliency map generated with ACAT, aiming to restore the normal anatomical structure based on the contextual information from the surrounding regions. Meanwhile, the regions of the image that do not contain any pathological elements are restored to their original appearance using DDIM sampling. Throughout the sampling process, these two components are fused together to ensure a seamless and realistic transition between the edited and unedited parts, resulting in a final image $\bm{\hat{x}}_0$ with a visually coherent and natural appearance} \label{fig:flow}
\end{figure*}

Another clinical application in which the detection of the volume of a lesion plays an important role is stroke management. In particular, it is important in the prognostic decision, in the  selection process for acute treatment~\cite{marks1999evaluation}, and in anticipating complications~\cite{mori2001aggressive}. 
Estimates of the tissue at risk and of the ischemic core are usually derived using Computed tomographic perfusion (CTP), perfusion-weighted imaging (PWI) or MRI diffusion-weighted imaging (DWI)~\cite{powers2019guidelines}. Software packages that automatically compute these estimates from perfusion imaging were also developed to facilitate clinical decisions about stroke treatment~\cite{mokli2019computer}. However, Computed Tomography (CT) scans are the most commonly used tool in stroke imaging, due to being inexpensive, efficient and widely available~\cite{mokli2019computer}. Consequently, quantitative measurements of the signs of infarction from CT scans, while more difficult to perform than on perfusion images, would be helpful in clinical practice. 

For these reasons, we propose a weakly-supervised method that is able to automatically segment brain tumours in MRI images and stroke lesions in CT scans. In particular, we generate anomaly maps without using pixel-level annotations of the anomalies, but using exclusively image-level labels (that are needed only at training time). The same methodology could also be applied to other pixel-wise anomaly detection tasks in medical images.

Radiologists' perception of machine learning tools varies from acceptance and enthusiasm to skepticism~\cite{pakdemirli2019artificial}. Providing simple anomaly maps could be negatively received by highly trained radiologists, who could consider it demeaning to their expertise~\cite{pakdemirli2019perception}. For this reason, in our approach we remove the lesions from pathological images and generate anomaly maps based on the difference between the original image and its normal-looking version. The healthy version of the image could be provided in place, or in addition, to the anomaly map, in order to better engage with clinicians and allow them to use their own inference to detect abnormalities. Indeed, radiologists usually detect deviations from a mental representation of the normal image~\cite{kundel1978visual}. Having a representation of the inner workings of the automatic image segmentation tool could also increase clinicians' trust in the model~\cite{arun2021assessing}.   Moreover, comparing normal and abnormal images is a common practice when teaching radiologists~\cite{xie2020chexplain}. Since normal anatomy can vary a lot, it is important for trainees to be exposed to a high number of healthy images~\cite{pakdemirli2019artificial}. However, the majority of teaching files are skewed towards pathological samples~\cite{boutis2016interpretation}. Therefore, by transforming abnormal examples to match normal anatomy, we could prevent this data imbalance and aid more effective training of radiologists. 

Previous work has employed autoencoders~\cite{zimmerer2018context,chen2018unsupervised,seebock2016identifying} or GANs~\cite{schlegl2017unsupervised,keshavamurthy2021weakly,siddiquee2019learning} trained on healthy samples to map diseased images to their corresponding normal version. However, autoencoders do not always produce sharp images and do not guarantee a correct mapping to the healthy version. On the other hand, GAN training can sometimes be unstable, depend on many hyperparameters and generate poor samples~\cite{shmelkov2018good}. For this reason, our approach is based on diffusion models, a class of generative models that have recently risen in popularity in the computer vision community due to their remarkable capabilities. They have been shown to achieve sample quality that is superior to the previous state-of-the-art GANs~\cite{dhariwal2021diffusion}. 

In~\cite{wolleb2022diffusion}, the authors employed diffusion models and classifier guidance~\cite{dhariwal2021diffusion} to recover the normal anatomy. However, the gradients that are needed to guide the sampling process have to be computed from a classifier trained on noised samples. This classifier often produces unreliable predictions, since in medical imaging the class of a sample is often determined by small details, that can be lost after only a few noising steps. For this reason, with this approach, we are not guaranteed to preserve the original structure of the sample and many details of the normal tissue can be modified.

A recent study~\cite{fontanella2023acat} has introduced Adversarial Counterfactual Attention (ACAT), an approach for mapping diseased images to their healthy counterparts and identifying Regions of Interest (ROIs). In particular, to generate counterfactual examples, the authors utilise an autoencoder and a classifier trained separately to reconstruct and classify images respectively. Specifically, they determine the minimal shift in the latent space of the autoencoder that transitions the input image towards the desired target class, as determined by the classifier's output. The authors extensively compared various counterfactual and gradient-based approaches for generating attribution maps to identify diseases in brain and lung CT scans. They demonstrated that their approach for generating saliency maps achieved the highest score in localizing the lesion location among six potential regions in brain CT scans. Moreover, it yielded the best Intersection over Union (IoU) and Dice score on lung CT scans.

\begin{figure*}[ht]
            \centering
            \begin{subfigure}[t]{0.18\linewidth}
                \includegraphics[width=\textwidth]{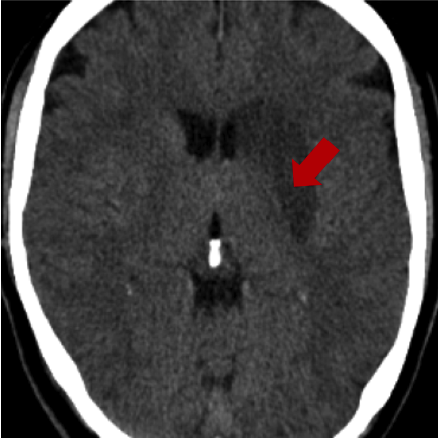}
                \caption{Original Image}
            \end{subfigure} \\
        \begin{subfigure}{0.118\textwidth}
            \includegraphics[width=\linewidth]{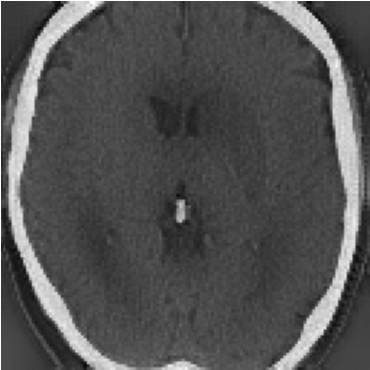}
        \end{subfigure} 
        \begin{subfigure}{0.118\textwidth}
            \includegraphics[width=\linewidth]{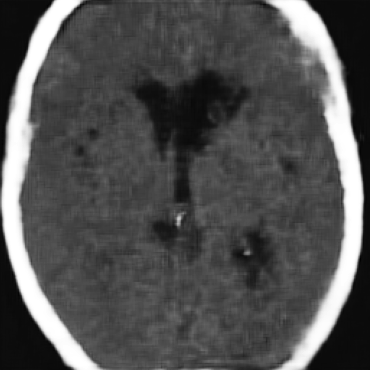}
        \end{subfigure} 
                \begin{subfigure}{0.118\textwidth}
            \includegraphics[width=\linewidth]{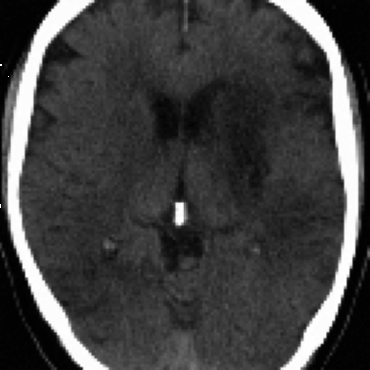}
        \end{subfigure} 
            \begin{subfigure}{0.118\textwidth}
            \includegraphics[width=\linewidth]{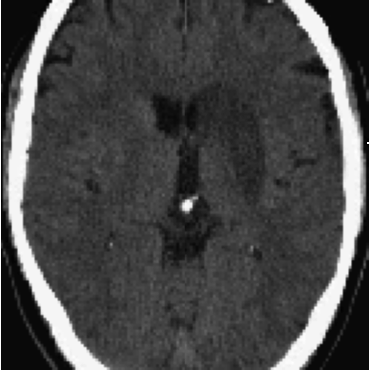}
        \end{subfigure} 
        \begin{subfigure}{0.118\textwidth}
            \includegraphics[width=\linewidth]{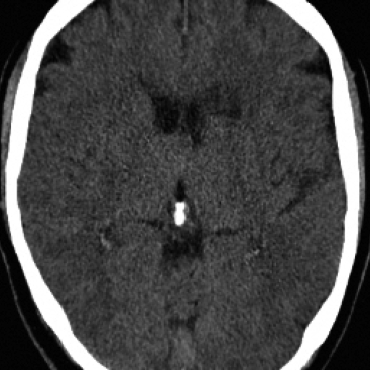}
        \end{subfigure} 
        \begin{subfigure}{0.118\textwidth}
            \includegraphics[width=\linewidth]{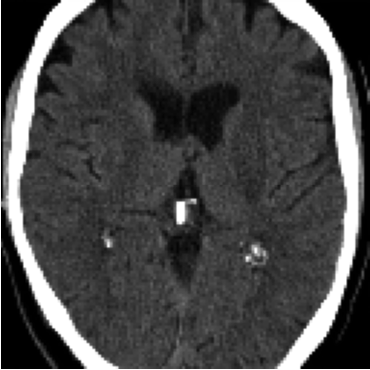}
        \end{subfigure} 
        \begin{subfigure}{0.118\textwidth}
            \includegraphics[width=\linewidth]{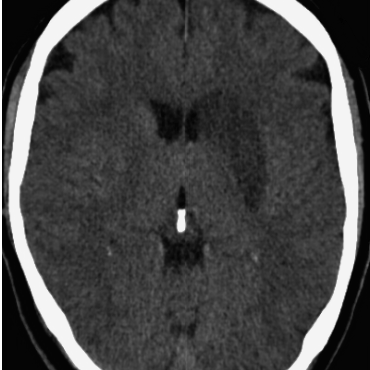}
        \end{subfigure} 
        \begin{subfigure}{0.118\textwidth}
            \includegraphics[width=\linewidth]{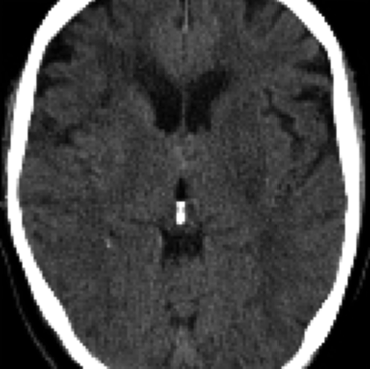}
        \end{subfigure}\\
        
        \begin{subfigure}[t]{0.118\textwidth}
            \includegraphics[width=\linewidth]{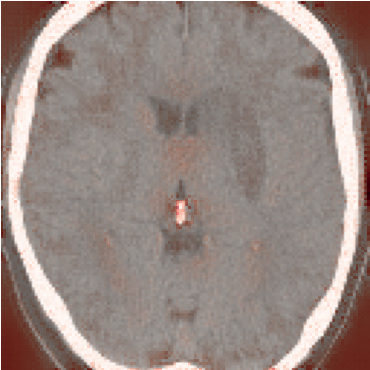}
            \caption{DenoisingAE}
        \end{subfigure} 
        \begin{subfigure}[t]{0.118\textwidth}
            \includegraphics[width=\linewidth]{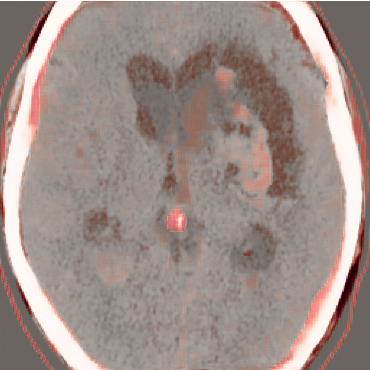}
            \caption{f-AnoGan}
        \end{subfigure} 
        \begin{subfigure}[t]{0.118\textwidth}
            \includegraphics[width=\linewidth]{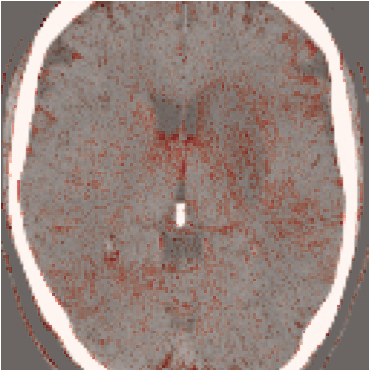}
            \caption{AnoDDPM}
        \end{subfigure} 
        \begin{subfigure}[t]{0.118\textwidth}
            \includegraphics[width=\linewidth]{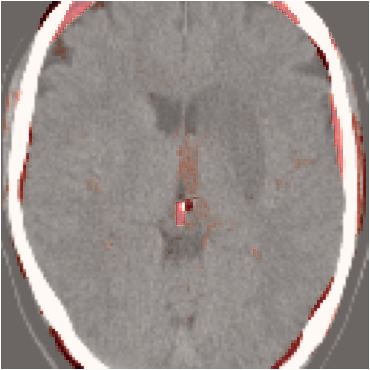}
            \caption{AutoDDPM}
        \end{subfigure} 
        \begin{subfigure}[t]{0.118\textwidth}
            \includegraphics[width=\linewidth]{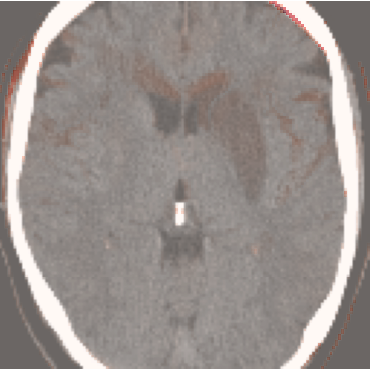}
            \caption{CG}
        \end{subfigure} 
         \begin{subfigure}[t]{0.118\textwidth}
            \includegraphics[width=\linewidth]{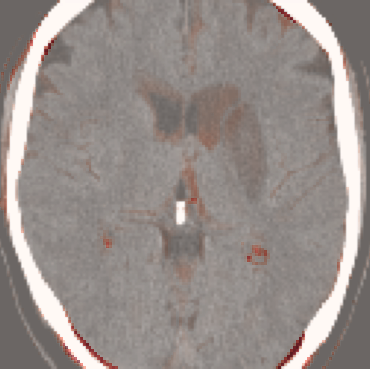}
            \caption{CFG}
        \end{subfigure} 
        \begin{subfigure}[t]{0.118\textwidth}
            \includegraphics[width=\linewidth]{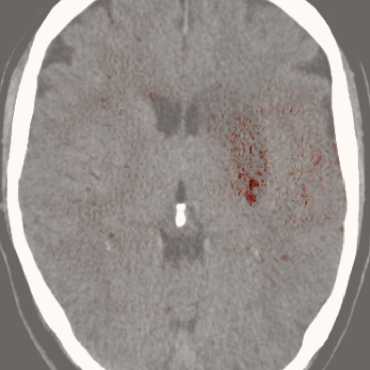}
            \caption{ACAT}
        \end{subfigure}       
        \begin{subfigure}[t]{0.118\textwidth}
            \includegraphics[width=\linewidth]{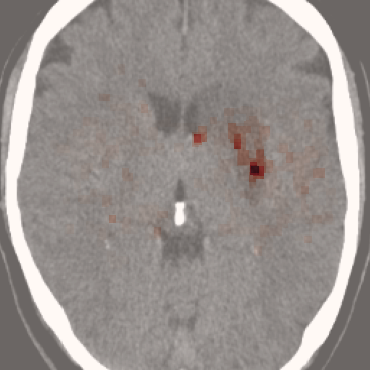}
            \caption{Dif-fuse (Ours)} 
        \end{subfigure} 
    \caption{Original image from IST-3 (a) and healthy counterfactuals (second row) with corresponding anomaly maps (bottom row), obtained with DenoisingAE (b), f-AnoGAN (c), AnoDDPM (d), AutoDDPM (e), classifier guidance (f), classifier-free guidance (g), ACAT (h) and Dif-fuse (i). ACAT generates a reasonable anomaly map, but is not able to fully remove the lesion. Dif-fuse refines the anomaly map obtained with ACAT, while at the same time creating a credible counterfactual example. The other approaches introduce artifacts and/or identify the pathological area less correctly.}
    \label{fig.plotist}
\end{figure*}

While ACAT revolves around generating counterfactuals, its primary strength lies in accurately identifying pathological regions, which are subsequently employed in a classification pipeline. On the other hand, it falls short in producing credible counterfactual examples, an issue we aim to address in this study. 
An illustration of this phenomenon is depicted in Fig.~\ref{fig.plotist}, where we can observe how ACAT is able to generate a saliency map that approximately identifies the pathological region (e, bottom row). However, in the counterfactual example, the lesion remains visible (e, top row). In contrast, our approach not only refines the saliency map but also generates a counterfactual image where the pathology is completely eliminated (f).

In order to do so, we exploit the saliency maps obtained with ACAT to guide the image generation process of diffusion models. We first train a Denoising Diffusion Probabilistic Model (DDPM) ~\cite{ho2020denoising} on healthy samples and use a combination of DDPM and Denoising Diffusion Implicit Model (DDIM) ~\cite{song2020denoising} sampling to remove pathological areas from the images. In particular, we first map an abnormal image to its noised version by using the reversed sampling approach of DDIMs. Then, with DDPM sampling we modify the pathological area, identified by the saliency map obtained previously, to recover the normal structure, based on the surrounding anatomical context. The parts of the image without pathological elements are mapped back to their original appearance with DDIM sampling. We fuse these two components at each step of the sampling process, so that the final resulting image has a realistic appearance, with a smooth transition between edited and unedited parts. We refer to our method as Dif-fuse.

In summary, our main contributions are: 1) we introduce a novel dual sampling strategy for diffusion models that, without the need for lesion annotations, allows inpainting of ROIs identified by a segmentation mask while preserving the rest of the image. Our innovation lies in the approach to mixing the two components at each timestep, resulting in a smooth fusion between edited and unedited parts. This enables the generation of realistic counterfactual examples of medical images as well as anomaly maps of the pathological areas; 2) We compare our approach with existing weakly supervised methods for medical image segmentation on WMH and BraTS 2021 datasets, achieving the highest mean Dice and IoU scores among the methods considered on both datasets. Our approach also has comparable image quality, as measured by the Kernel Inception Distance (KID)~\cite{binkowski2018demystifying} on IST-3, to unconstrained (without masking) diffusion sampling methods, with the added advantage of more accurate anomaly segmentation.

\section{Related Work}

\subsection{Saliency Maps}
Saliency maps are frequently utilised by researchers to gain insights into the inner workings of neural networks. They aid in the interpretation of convolutional neural network (CNN) predictions by emphasizing the significance of pixels in determining model outcomes. 
In~\cite{simonyan2013deep}, the authors employed the gradient of the target class's score relative to the input image, while the Guided Backpropagation method~\cite{springenberg2014striving} backpropagates solely positive gradients. The Integrated Gradient method~\cite{sundararajan2017axiomatic} integrates gradients between the input image and a baseline black image. Smilkov et al.~\cite{smilkov2017smoothgrad} introduced SmoothGrad, which involves smoothing the gradients using a Gaussian kernel. 

Grad-CAM~\cite{selvaraju2017grad}, which builds on the Class Activation Mapping (CAM) approach~\cite{zhou2016learning}, employs the gradients of the target class's score with respect to the feature activations of the final convolutional layer to determine the importance of spatial locations. Given the popularity of this approach, modifications and improvements were later proposed in several papers.
For example, Grad-CAM++~\cite{chattopadhay2018grad} introduced pixel-wise weighting of the gradients of the output with respect to a particular spatial position in the final convolutional layer. In this way, it is able to obtain a measure of the importance of each pixel in a feature map for the classification outcome.
On the other hand, Score-CAM~\cite{wang2020score} takes a different approach by eliminating the dependency on gradients altogether. Instead, it calculates the weights of each activation map through a forward passing score for the target class.

\subsection{Counterfactual Explanations}

\begin{figure*}[htb]
\hspace{\fill}
\begin{subfigure}{0.32\textwidth}
\includegraphics[width=1\textwidth]{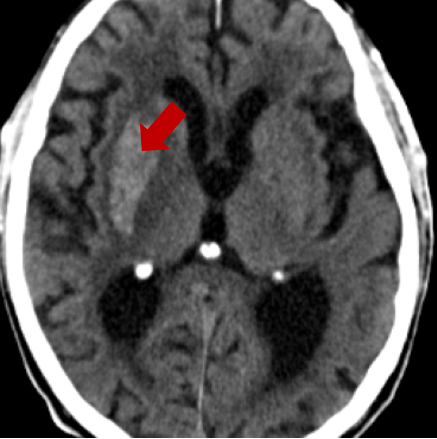}
\caption{}
\end{subfigure}\hspace{\fill}
\begin{subfigure}{0.32\textwidth}
\includegraphics[width=1\textwidth]{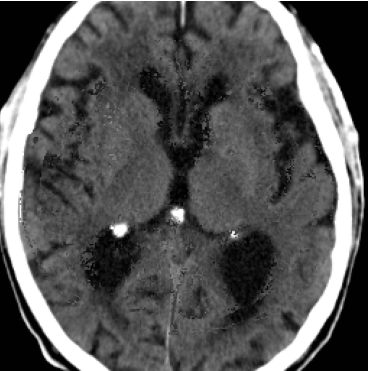}

\caption{}
\end{subfigure}\hspace{\fill}
\caption{Input image from IST-3 (a) and normal image generated by applying the mask only at the end of the sampling process (b). We can observe that (b) presents some artifacts and does not have a smooth transition between edited and unedited parts.}
\label{fig.baseline}
\end{figure*}

Previous work has demonstrated that gradient-based methods for generating saliency maps have limitations. In particular, they are not reliable in identifying critical regions in medical images, as highlighted by \cite{eitel2019testing} and \cite{arun2021assessing}, and have been shown to be independent of model parameters and training data, as demonstrated by \cite{adebayo2018sanity} and \cite{arun2021assessing}. As a result, techniques for visual explanations based on counterfactual examples have been developed. These methods typically involve learning a mapping between images of multiple classes to emphasize the relevant areas for each image's respective class. The mapping is typically modeled using a CNN and trained using a GAN. In particular, Baumgartner et al.~\cite{baumgartner2018visual} employed a Wasserstein GAN~\cite{arjovsky2017wasserstein}. Schutte et al.~\cite{schutte2021using} trained a StyleGAN2~\cite{karras2020analyzing} and looked for the minimal modification in the latent space that keeps the image as close as possible to the original one, but changes the class prediction. In~\cite{singla2021explaining}, the authors used a conditional Generative Adversarial Network (cGAN) to create a series of perturbed images that gradually display the transition between positive and negative class.

Cohen et al.~\cite{cohen2021gifsplanation} proposed the latent shift method. Their approach involves training an autoencoder and a classifier as separate components: the autoencoder is responsible for reconstructing images, while the classifier focuses on image classification. Subsequently, the input images undergo perturbations in the latent space of the autoencoder, resulting in $\lambda$-shifted variations of the original image. These variations modify the probability of a particular class of interest, as determined by the classifier's output. 

In a recent study, \cite{fontanella2023acat} observed that the single-step optimisation procedure employed in the latent shift method is sometimes unable to correctly generate counterfactuals. They also note that the generated samples in the aforementioned method may deviate significantly from the original image and fail to remain on the data manifold. To address these challenges, the authors propose ACAT, an approach that enhances the optimization procedure by incorporating small progressive shifts in the latent space instead of a single-step shift of size $\lambda$ from the input image. In this way, the probability of the class of interest converges smoothly to the target value.  Additionally, they introduce a regularization term to restrict the movement in latent space and ensure that the observed changes can be attributed to the class shift, while preserving the important characteristics of the image.

\subsection{Anomaly Detection}
The detection of disease markers in medical images is an important component for diagnosing disease and monitoring its progression. However, pixel-wise annotations are expensive to collect and often unavailable. For this reason, unsupervised or weakly-supervised anomaly detection has gained significant interest in the research community.
The most popular approaches involve autoencoders, GANs, or more recently, diffusion models.

\begin{figure*}[ht]
\centering

\begin{subfigure}{0.118\textwidth}
\includegraphics[width=1\textwidth]{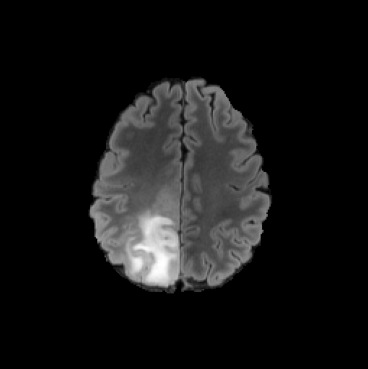}
\end{subfigure}
\begin{subfigure}{0.118\textwidth}
\includegraphics[width=1\textwidth]{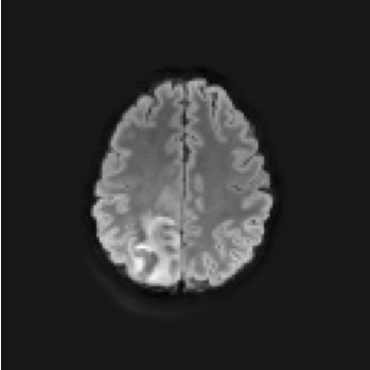}
\end{subfigure}
\begin{subfigure}{0.118\textwidth}
\includegraphics[width=1\textwidth]{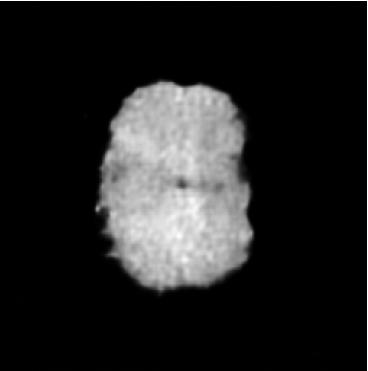}
\end{subfigure}
\begin{subfigure}{0.118\textwidth}
\includegraphics[width=1\textwidth]{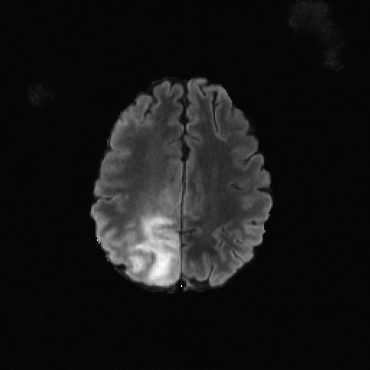}
\end{subfigure}
\begin{subfigure}{0.118\textwidth}
\includegraphics[width=1\textwidth]{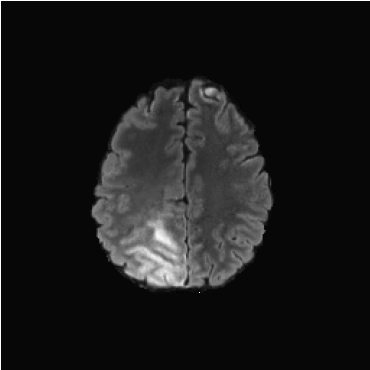}
\end{subfigure}
\begin{subfigure}{0.118\textwidth}
\includegraphics[width=1\textwidth]{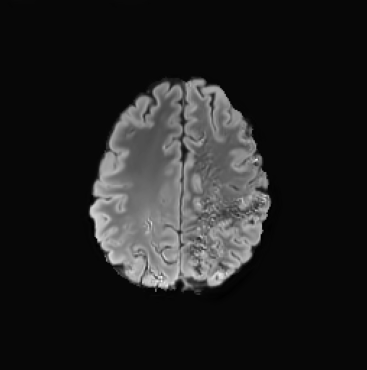}
\end{subfigure}
\begin{subfigure}{0.118\textwidth}
\includegraphics[width=1\textwidth]{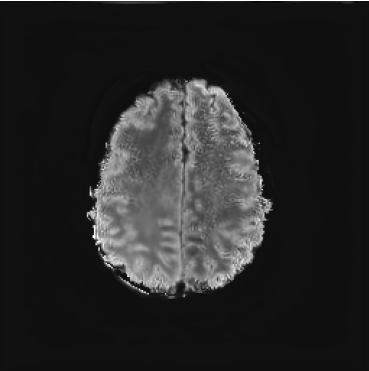}
\end{subfigure}
\begin{subfigure}{0.118\textwidth}
\includegraphics[width=1\textwidth]{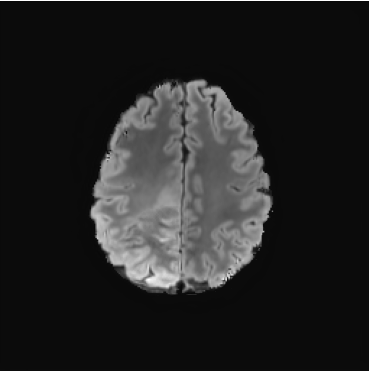}
\end{subfigure}\\

\begin{subfigure}[t]{0.118\textwidth}
\includegraphics[width=1\textwidth]{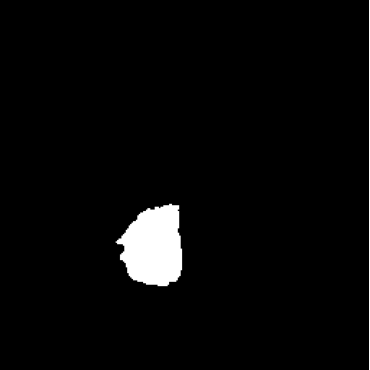}
\caption{Image}
\end{subfigure}
\begin{subfigure}[t]{0.118\textwidth}
\includegraphics[width=1\textwidth]{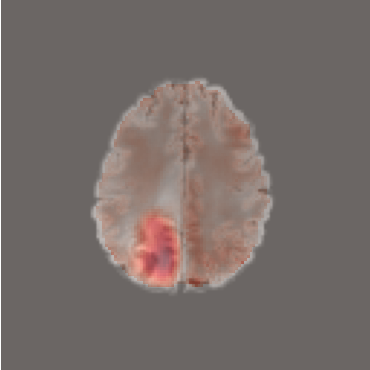}
\caption{DenoisingAE}
\end{subfigure}
\begin{subfigure}[t]{0.118\textwidth}
\includegraphics[width=1\textwidth]{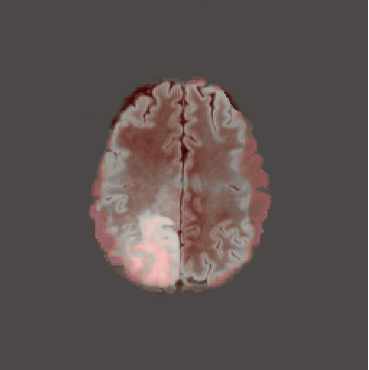}
\caption{f-AnoGan}
\end{subfigure}
\begin{subfigure}[t]{0.118\textwidth}
\includegraphics[width=1\textwidth]{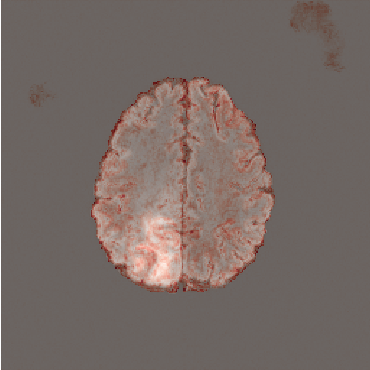}
\caption{AnoDDPM}
\end{subfigure}
\begin{subfigure}[t]{0.118\textwidth}
\includegraphics[width=1\textwidth]{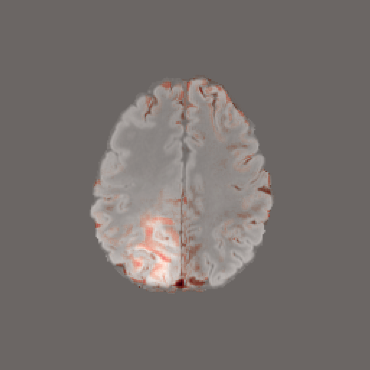}
\caption{AutoDDPM}
\end{subfigure}
\begin{subfigure}[t]{0.118\textwidth}
\includegraphics[width=1\textwidth]{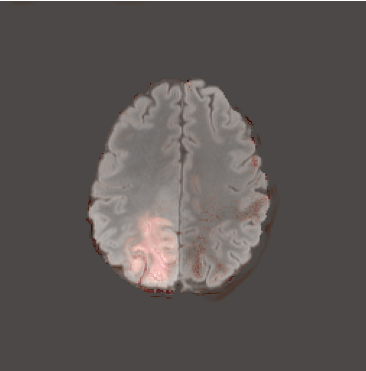}
\caption{CG}
\end{subfigure}
\begin{subfigure}[t]{0.118\textwidth}
\includegraphics[width=1\textwidth]{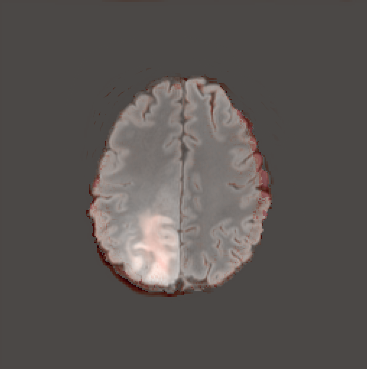}
\caption{CFG}
\end{subfigure}
\begin{subfigure}[t]{0.118\textwidth}
\includegraphics[width=1\textwidth]{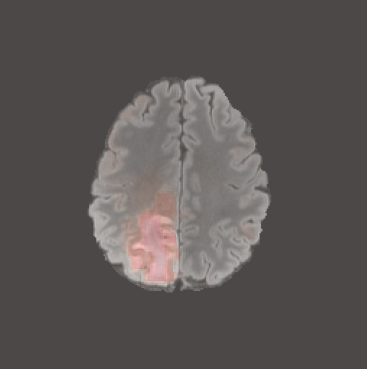}
\caption{Dif-fuse}
\end{subfigure}

\caption{Original image from BraTS 2021 with ground truth segmentation mask (a) and healthy images (top row) with corresponding anomaly maps (bottom row), obtained with DenoisingAE (b), f-AnoGan (c), AnoDDPM (d), AutoDDPM (e), classifier guidance (f), classifier-free guidance (g) and with Dif-fuse (h). f-Ano GAN falls short in generating believable counterfactuals, whereas the other approaches yield higher-quality results. However, DenoisingAE, AnoDDPM and AutoDDPM do not fully remove the lesion, while the counterfactuals generated with CG and CFG exhibit some artifacts.}
\label{fig.plot}
\end{figure*}

A common approach when employing autoencoders is to train them to reconstruct data from healthy subjects~\cite{zimmerer2018context,chen2018unsupervised,seebock2016identifying, kascenas2022denoising}. At test time, diseased images are mapped to the training distribution of healthy patients. The difference between the diseased input and the healthy output is the anomaly map. 

Schlegl et al.~\cite{schlegl2019f} propose f-AnoGAN, which follows a similar approach but with GANs. In particular, they train a generative model and a discriminator to distinguish between generated and real data. They also propose a mapping scheme to evaluate new data at test time and identify anomalous regions.
Other authors employed weakly supervised GANs, trained on both healthy and diseased images. In~\cite{keshavamurthy2021weakly}, the authors trained a Wasserstein GAN on unpaired chest x-rays images and learned to map diseased images to healthy ones. 

Diffusion models were employed in~\cite{wolleb2022diffusion}, where the authors first trained a probabilistic diffusion model on both diseased and healthy images, together with a binary classifier trained on noised samples. Then, they employed deterministic sampling from DDIM and classifier guidance~\cite{dhariwal2021diffusion} to map a diseased image into a healthy one. Another technique to guide diffusion models, proposed by \cite{ho2022classifier},  is classifier-free guidance. During training, the label of a class-conditional diffusion model is replaced with a null label with a fixed probability. During sampling, to guide the generation process, the output of the model is extrapolated further in the direction of the desired label and away from the null label.

An issue of these guidance-based approaches is that they either rely on a binary classifier trained on noised samples (in the case of classifier guidance) or an implicit classifier for noised samples through joint training of conditional and unconditional models (in the case of classifier-free guidance). While these approaches can work well for natural images, in our experiments they proved less effective for medical images, where adding noise can quickly erase most class-specific information, making the guidance unreliable.

In concurrent work, Bercea et al.~\cite{bercea2023reversing} use masks to inpaint pathological areas, applying the mask after generating the normal image with a GAN. This approach does not ensure smooth transitions at the mask boundaries. In follow-up work by some of the same authors~\cite{bercea2023mask}, an approach that employs masking, stitching, and resampling with a diffusion model is proposed. 
In particular, they obtain the mask for the pathological area directly using the diffusion model trained on normal samples, by noising and then denoising the pathological image to obtain $X_{rec}$, before computing the residual between $X_{rec}$ and the original image to obtain the mask. Then, they mask the original image, obtaining $X_m$, re-noise $X_{rec}$ to timestep $t$ to obtain $X_{rec}^t$, apply a sampling step with the diffusion model to compute  $X_{rec}^{t-1}$  and noise $X_m$  to timestep $t-1$. Since $X_{rec}^{t-1}$ and $X_m^{t-1}$  are not fully compatible with each other as they were obtained independently, the authors introduce some additional harmonization steps to blend the two components better. This process is repeated at each timestep. 

In our approach, we side-step the blending issue by applying DDIM inversion to the entire image. Then, at each step we apply both DDIM (for reconstructing the areas outside of the mask) and DDPM (for removing the pathology from the area inside the mask) on the same noised image, before applying the masking operation. This has the advantage that both processes are applied by always considering the entire anatomical context and the harmonization and blending process is naturally carried out in a gradual way during the entire sampling process, without needing to add additional computational overhead for explicit harmonization procedures.  

Our approach and the work by Bercea et al. ~\cite{bercea2023mask} also have different tradeoffs. While their method performs masking in an unsupervised way, which avoids introducing bias on expected anomaly distributions, our method relies on a classifier trained in a weakly supervised manner. Bercea’s approach may lead to suboptimal, fragmented masks that could still include areas of pathology. Our method, on the other hand, aims to provide more consistent and comprehensive masking of anomalous regions. Both approaches have their merits, and the choice between them may depend on the specific application and data availability. We recognise that the behavior of our classifier outside known classes requires further investigation. Further research could explore the generalisability of our classifier-based approach to a wider range of diseases, including rare or previously unseen conditions.
While our method requires an external model to compute the initial saliency maps, \cite{bercea2023mask} requires tuning several critical hyperparameters: the amount of noise for computing the mask, the masking threshold, the noise level for inpainting, and the number of resampling steps. Finding the right combination of these hyperparameters can be computationally expensive and complex. In contrast, we aimed to reduce the hyperparameter space for easier tuning. Our method primarily depends on the noising amount and masking threshold, simplifying the process of hyperparameter optimization.

\section{Methods}

ACAT addresses the limitations of the latent shift method in generating attribution maps; however, the counterfactual examples obtained through ACAT are not entirely satisfactory. In other words, ACAT is able to identify where an image should be modified, but not exactly how to modify it to obtain a credible counterfactual. This limitation is understandable since the primary focus of their paper was to utilise these saliency maps within a classification pipeline rather than generating precise counterfactuals. 

In our work, we aim to tackle this challenge by proposing a two-step approach. First, we employ ACAT to obtain initial saliency maps, which provide a rough identification of the regions requiring modification. Then, we introduce a novel sampling technique from diffusion models that enables targeted modifications to these regions while preserving the remainder of the image unchanged. By fusing both components at each timestep, we achieve a seamless transition between the edited and unedited parts, resulting in a realistic output. By considering the difference between the counterfactual example and the original image, we can also obtain the final anomaly map. 

We observe that our sampling approach not only generates highly realistic counterfactuals but also enhances the initial saliency maps obtained in the first step using ACAT. This is possible because the selected regions may not undergo complete modification by the diffusion model, allowing for the preservation of healthy anatomical features identified in the initial attribution maps. A visual representation of our approach is presented in Fig.~\ref{fig:flow}.

In the next sections, we first give a brief overview of diffusion models, before introducing our sampling technique to generate credible counterfactuals and obtain pixel-wise anomaly maps of pathological areas in medical images.

\subsection{Diffusion Models}

A diffusion model is defined by a forward process that gradually adds noise to data starting from $\bm{x}_0 \sim q(\bm{x}_0)$ over $T$ timesteps~\cite{ho2020denoising}:
\begin{equation}
\label{Eq.1}
q(\bm{x}_{1:T}|\bm{x}_0) = \prod_{t=1}^{T} q(\bm{x}_t|\bm{x}_{t-1})
\end{equation}
with $q(\bm{x}_t|\bm{x}_{t-1}) =  \mathcal{N}(\bm{x}_t; \sqrt{1-\beta_t}\bm{x}_{t-1}, \beta_t \bm{I})$

and a backward process: $p_{\theta}(\bm{x}_0) = \int p_\theta(\bm{x}_{0:T})d\bm{x}_{1:T}$, where:

\begin{equation}
\begin{aligned}
\label{Eq.mu}
p_{\theta}(\bm{x}_{0:T}) &= p(\bm{x}_T)\prod_{t=1}^{T} p_{\theta}(\bm{x}_{t-1}|\bm{x}_t), \; \;\\
p_{\theta}(\bm{x}_{t-1}|\bm{x}_t) &=  \mathcal{N}(\bm{x}_{t-1};\bm{\mu}_{\theta}(\bm{x}_t, t), \bm{\Sigma}_{\theta}(\bm{x}_t, t))
\end{aligned}
\end{equation}

The parameters of the forward process $\beta_t$ are set so that $\bm{x}_T$ is distributed approximately as a standard normal distribution and therefore $p(\bm{x}_T)$ is set to a standard normal prior too. We can train the backward process to match the distribution of the forward process by optimising the evidence lower bound (ELBO): $-L_\theta(\bm{x}_0) \le log (p_\theta (\bm{x}_0))$:

\begin{equation}
\begin{aligned}
\label{Eq.2}
L_{\theta}(\bm{x}_{0}) &=  \mathbb{E}_q[L_T(\bm{x}_0)\\
&+\sum_{t>1}D_{KL}(q(\bm{x}_{t-1}|\bm{x}_t, \bm{x}_0) || p_\theta (\bm{x}_{t-1}|\bm{x}_t))\\
&-log p_\theta(\bm{x}_0|\bm{x}_1)] 
\end{aligned}
\end{equation}
where $L_{T}(\bm{x}_{0}) = D_{KL}(q(\bm{x}_{T}|\bm{x}_0) || p(\bm{x}_{T}))$. 

The forward process posteriors $q(\bm{x}_{t-1}|\bm{x}_t, \bm{x}_0)$ and marginals $q(\bm{x}_{t}|\bm{x}_0)$ are Gaussian and the KL divergence can be calculated in closed form. Therefore, the diffusion model can be trained by taking stochastic gradient descent steps on random terms of~(\ref{Eq.2}). As noted in~\cite{ho2020denoising}, the noising process defined in~(\ref{Eq.1}) allows us to sample arbitrary steps of the latents, conditioned on ${x}_0$. With $\alpha_t:= 1-\beta_t$ and $\hat{\alpha}_t:=\prod_{s=0}^t \alpha_s$, we can write:
\begin{equation}
\label{Eq.3}
q(\bm{x}_{t}|\bm{x}_0) =  \mathcal{N}(\bm{x}_t; \sqrt{\alpha_t}\bm{x}_{0}, (1-\hat{\alpha}_t)\bm{I}).
\end{equation}
Therefore:
\begin{equation}
\label{Eq.noising}
\bm{x}_t = \sqrt{\alpha_t}\bm{x}_{0}+\sqrt{(1-\hat{\alpha}_t)} \bm{\epsilon}, 
\end{equation}
with $\bm{\epsilon} \sim \mathcal{N}(\bm{0},\bm{I})$.

There are many ways to parametrise $\bm{\mu}_\theta (\bm{x}_t, t)$ (\ref{Eq.mu}) in the prior. For example, we could predict $\bm{\mu}_\theta (\bm{x_t}, t)$ with a neural network. Alternatively, we could predict $\bm{x}_0$ and use it to compute $\bm{\mu}_\theta (\bm{x_t}, t)$. The network could also be used to predict the noise $\bm{\epsilon}$. In~\cite{ho2020denoising}, the authors found that this option produced the best sample quality and introduced the reweighted loss function:
\begin{equation}
\label{Eq.4}
L_{\textrm{simple}} = \mathbb{E}_{t,\bm{x}_0, \epsilon} [|| \bm{\epsilon} - \bm{\epsilon}_\theta(\bm{x}_t, t) ||^2]
\end{equation}

After training the diffusion model, given $\bm{X}_T$ sampled from $\mathcal{N}(\bm{0},\bm{I})$, we can generate a new sample $\bm{x}_0$, by recursively applying the sampling scheme:
\begin{equation}
\begin{aligned}
\label{Eq.sampling}
\bm{x}_{t-1} &= \sqrt{\hat{\alpha}_{t-1}}\left(\frac{\bm{x}_t-\sqrt{(1-\hat{\alpha}_t)} \bm{\epsilon}_\theta(\bm{x}_t, t)}{\sqrt{\hat{\alpha}_t}}\right) \\
&+ \sqrt{1 - \hat{\alpha}_{t-1} - \sigma^2_t} \bm{\epsilon}_\theta(\bm{x}_t, t) +\sigma_t \bm{\epsilon}
\end{aligned}
\end{equation}
IN DDPMs:
\begin{equation*}
\sigma_t = \sqrt{(1 - \hat{\alpha}_{t-1})/(1 - \hat{\alpha}_{t})} \sqrt{1 - \hat{\alpha}_{t}/\hat{\alpha}_{t-1})}.
\end{equation*}
IN DDIMs, the stochastic component is removed by setting $\sigma_t = 0$ and the sampling process becomes deterministic.

\subsection{Dif-fuse}

\begin{algorithm}[t]
\caption{Dif-fuse }\label{alg:dif}
\begin{algorithmic}
\STATE {\bfseries Input:} Image $\bm{x}_0$, noise amount $K$, threshold value $\tau$
\STATE {\bfseries Output:} Counterfactual image $\hat{\bm{x}}_{0}$, anomaly map $\delta$
\STATE Compute saliency map of $\bm{x}_0$ with ACAT
\STATE Smooth the saliency map with a $5 \times 5$ kernel and binarise it using the threshold $\tau$, to obtain the mask $\bm{m}$
\FORALL{t in $[0,K-1]$} 
\STATE{$\bm{x}_{t+1} \gets \bm{x}_t+ \sqrt{\hat{\alpha}_{t+1}}\Biggl[\Biggl(\sqrt{\frac{1}{\hat{\alpha}_t}}-\sqrt{\frac{1}{\hat{\alpha}_{t+1}}}\Biggl)\bm{x}_t 
+ \left( \sqrt{\frac{1}{\hat{\alpha}_{t+1}}-1} - \sqrt{\frac{1}{\hat{\alpha}_{t}}-1} \right) \bm{\epsilon}_\theta(\bm{x}_t, t) \Biggl]$} 
\ENDFOR
\FORALL{t in $[K,1]$} 
\STATE{$\hat{\bm{x}}_{t-1} \gets \bm{x}^{DDPM}_{t-1}\odot \bm{m}+\bm{x}^{DDIM}_{t-1}\odot(1-\bm{m})$} 
\ENDFOR
\STATE $d = |\bm{x}_0 - \hat{\bm{x}}_{0} | $
\STATE Apply erosion followed by dilation
with a $5\times5$ kernel to d
\STATE Apply dilation followed by erosion to obtain the final anomaly map $\delta$
\RETURN $\hat{\bm{x}}_{0}$, $\delta$
\end{algorithmic}
\end{algorithm}

\begin{figure*}[t]
\centering
\includegraphics[width=0.6\textwidth]{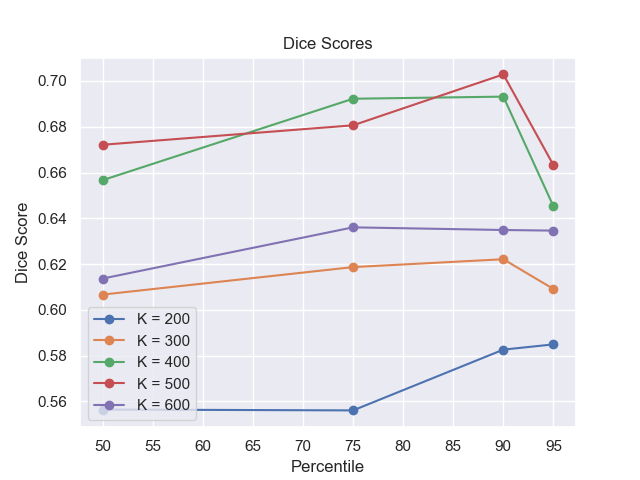}
\caption{Dice scores obtained on the validation dataset with different combinations of thresholding percentiles to binarise the saliency maps and noise amounts K. We obtain the best result with $K=500$ and pixels in the $90^{th}$ percentile of the saliency maps.} \label{fig:hyp}
\end{figure*}

In our approach, we employ a DDPM trained on healthy samples and saliency maps obtained from adversarially generated counterfactual examples as in ACAT~\cite{fontanella2023acat}. We chose ACAT as it showed superior performance in the identification of pathological areas in brain and lung CT scans. However, in principle, saliency maps may also be generated with any other approach. Given a diseased image $\bm{x}_0$, we first select a noise amount $K \in [0,T]$ and map the image to its noised version $\bm{x}_K$ with the inverse DDIM sampling scheme proposed in~\cite{song2020denoising}:
\begin{equation}
\begin{aligned}
\label{inv_DDIM}
\bm{x}_{t+1} &= \bm{x}_t+ \sqrt{\hat{\alpha}_{t+1}}\Biggl[\Biggl(\sqrt{\frac{1}{\hat{\alpha}_t}}-\sqrt{\frac{1}{\hat{\alpha}_{t+1}}}\Biggl)\bm{x}_t \\
&+ \left( \sqrt{\frac{1}{\hat{\alpha}_{t+1}}-1} - \sqrt{\frac{1}{\hat{\alpha}_{t}}-1} \right) \bm{\epsilon}_\theta(\bm{x}_t, t) \Biggl]
\end{aligned}
\end{equation}
We then smooth the saliency map with a Gaussian kernel of size $5 \times 5$ to obtain a mask $\bm{m}$ that is more uniform and with fewer isolated pixels. We edit the diseased regions inside the mask with DDPM sampling. Since the diffusion model was trained on normal samples, these regions are mapped to a healthy appearance. The rest of the anatomy needs to be preserved and therefore we employ DDIM sampling for the areas outside of the mask, as in~(\ref{Eq.sampling}), with $\sigma_t = 0$. In order to obtain a coherent result, we mix the masked part with the rest of the image at each sampling step. In other words, given $\hat{\bm{x}}_{t}$, we compute:
\begin{equation}
\label{Eq.mix}
\hat{\bm{x}}_{t-1} = \bm{x}^{DDPM}_{t-1}\odot \bm{m}+\bm{x}^{DDIM}_{t-1}\odot(1-\bm{m})
\end{equation}
where $\odot$ is the Hadamard product.
In this way, the editing process is focused on the parts that were captured by the saliency map, preventing random changes to the structural characteristics of the scan. In fact, the DDIM sampling guarantees reconstruction of the parts that don't need to be edited. Moreover, changes to the pathological parts are performed by the DDPM considering the surrounding anatomical context. Our method is summarised in Algorithm ~\ref{alg:dif}.

When computing $\hat{\bm{x}}_{t-1}$ with~(\ref{Eq.mix}), the sum of the two components may not produce a perfectly coherent result. However, the incoherence is resolved by the next diffusion step, which fuses the two components better. This would not be the case if we simply computed $\hat{\bm{x}}_0$ with DDPM and then applied the mask only at the end of the sampling process. An illustration of this effect is presented in Fig.~\ref{fig.baseline}, where we can observe how the normal image, generated by applying the mask solely at the conclusion of the sampling process (b), exhibits some artifacts and lacks a seamless transition between the edited and unedited regions.

In this way, we are able to obtain a normal version of the given pathological image. In order to obtain an anomaly map, we first compute the difference between the original and the generated image and then apply erosion followed by dilation with a $5 \times 5$ kernel to the resulting map, in order to remove noise, and finally dilation followed by erosion, with the same kernel, to close small holes in the map.

\subsection{Training details}

The diffusion model is trained for 60,000 iterations, with a batch size of 10, using the loss proposed in ~\cite{nichol2021improved} and the AdamW optimiser, with learning rate of 1e-4, $\beta_1 = 0.9$, $\beta_2= 0.999$ and weight decay coefficient of 0.05. We used an EMA rate of 0.99 and a noise schedule as in \cite{ho2020denoising}, setting the forward process variances to constants that increase linearly from  $10^{-4}$ in the first step to $0.02$ in the last one. Training takes around two days on one NVIDIA A100 GPU. We employed 1000 sampling steps and a U-Net architecture with 128 channels in the first layer and attention heads at resolutions 8,16,32. The U-Net model employs a sequence of residual layers and downsampling convolutions, followed by another sequence of residual layers with upsampling convolutions. These layers are connected through skip connections, linking layers with the same spatial size. In particular, we used two residual blocks per resolution.

\begin{figure*}[htb]
\centering
\hspace{1cm}
\begin{subfigure}{0.2\textwidth}
\includegraphics[width=1\textwidth]{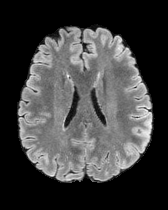}\\
\end{subfigure}\hspace{\fill}
\begin{subfigure}{0.2\textwidth}
\includegraphics[width=1\textwidth]{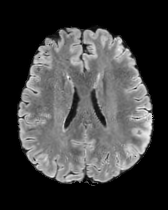}\\
\end{subfigure}\hspace{\fill}
\begin{subfigure}{0.2\textwidth}
\includegraphics[width=1\textwidth]{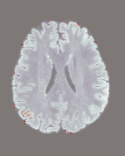}\\
\end{subfigure}\hspace{\fill}

\hspace{1cm}
\begin{subfigure}{0.2\textwidth}
\includegraphics[width=1\textwidth]{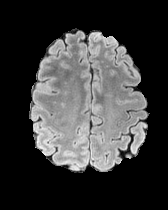}
\caption{}
\end{subfigure}\hspace{\fill}
\begin{subfigure}{0.2\textwidth}
\includegraphics[width=1\textwidth]{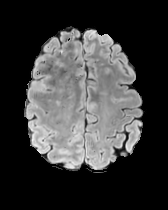}
\caption{}
\end{subfigure}\hspace{\fill}
\begin{subfigure}{0.2\textwidth}
\includegraphics[width=1\textwidth]{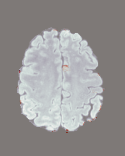}
\caption{}
\end{subfigure}\hspace{\fill}
\caption{Healthy input images from BraTS 2021 (a), images generated with Dif-fuse (b) and anomaly maps (c). We can observe that our approach obtains a good reconstruction of healthy samples. }
\label{fig.healthy}
\end{figure*}

\section{Experiments}

\subsection{Data}

We performed our experiments on IST-3~\cite{sandercock2011international},  BraTS 2021~\cite{baid2021rsna} and the White Matter Hyperintensity (WMH)~\cite{kuijf2019standardized} datasets.

IST-3 is a randomised-controlled trial that collected brain imaging data, primarily CT scans, from 3035 patients exhibiting stroke symptoms. The scans were conducted at two time points: immediately after the patients' hospital admission and again between 24-48 hours later. Radiologists involved in the trial assessed the presence or absence of early ischemic signs and recorded the location of any identified lesions for positive scans.
In our analysis, we considered a total of 5681 scans, $46.31\% $ of which were classified as negative (no lesion), while the remaining scans were positive. In particular, We considered 11 slices for each scan and resized each slice to $256 \times 256$. For more detailed information about the trial protocol, data collection, and the data use agreement, please refer to the following URL: \href{https://datashare.ed.ac.uk/handle/10283/1931}{IST-3 information{\footnote{\url{https://datashare.ed.ac.uk/handle/10283/1931}.}}}. 

BraTS 2021 includes data that was collected for the Brain Tumor Segmentation (BraTS) challenge. This dataset consists of pre-operative baseline multi-parametric magnetic resonance imaging (mpMRI) scans obtained with different clinical protocols and various scanners from multiple institutions. The primary objective of the challenge is to evaluate and compare advanced techniques for segmenting different sub-regions of intrinsically heterogeneous brain glioblastomas in mpMRI scans. It includes scans in four modalities (FLAIR, T1, T1 weighted and T2). In particular, we considered the publicly available BraTS 2021 training dataset, containing scans from 1251 patients. Each scan has 155 slices. However, we removed the top and bottom 25 slices, since they have minimal content, and any other empty ones, before zero padding the remaining to $256 \times 256$ (from the original dimension of $240 \times 240$). In the end, we are left with 131,164 slices, of which 79,113 are positive. Additional information on the dataset can be found here: \href{http://braintumorsegmentation.org/}{BraTS 2021 information\footnote{\url{http://braintumorsegmentation.org/}.}}. 

WMH was collected for the White Matter Hyperintensity segmentation challenge. We employed data from the test set, which is composed of 110 scans from five MR scanners of FLAIR and T1 modalities. We center-cropped and resized each slice to $256 \times 256$.

As annotations of lesions are not available in IST-3, we utilise this dataset to evaluate the quality of the generated images, rather than the segmentation accuracy. On the other hand, for the BraTS 2021 and WMH datasets, we have access to lesion annotations, enabling us to conduct quantitative analysis of the anomaly maps that we create. IST-3 and BraTS 2021 were divided into training, validation and test sets with a 70-15-15 split. On WMH, we evaluate the models trained on BraTS 2021 without further fine-tuning, to test their out-of-domain generalisation capabilities.

\subsection{Experimental Setup}
We compare our approach with competing weakly-supervised approaches employing autoencoders, GANs and diffusion models. In particular, we considered DenoisingAE~\cite{kascenas2022denoising}, following the implementation from the official repository\footnote{\url{https://github.com/AntanasKascenas/DenoisingAE}.}, f-Ano GAN~\cite{schlegl2019f}, in which we trained both WGAN and izi encoder for 500,000 iterations each, diffusion models with classifier guidance (CG) during sampling, following the implementation of~\cite{wolleb2022diffusion} with noise level $K=500$ and gradient scale $s=100$, classifier-free guidance (CFG)~\cite{ho2022classifier} with guidance scale $s'=3$ (which in our experiments obtained the best results). Additionally, we also evaluated AnoDDPM~\cite{wyatt2022anoddpm}\footnote{\url{https://github.com/Julian-Wyatt/AnoDDPM}.} and AutoDDPM~\cite{bercea2023mask}\footnote{\url{https://github.com/ci-ber/autoDDPM}.}. For the former, we observed on validation data that employing 100 noising steps achieves the best results, while for the latter we followed the hyperparameters of ~\cite{bercea2023mask} and set the masking threshold such that at most $5\%$ false positives are obtained, while tuning the final anomaly binarization threshold on validation data (the optimal threshold was found to be 0.1). 
As an ablation, we also consider the result obtained directly using the saliency maps obtained with ACAT, thresholded as in our approach, as anomaly maps and different combinations of DDIM and DDPM sampling for forward and backward sampling processes (without masking). In particular, we considered DDPM sampling from an image noised with the forward process of the diffusion model (called DDPM in the experiments), DDPM sampling starting from an image noised with DDIM inversion (DDIM-DDPM), DDIM sampling from an image noised with the forward process of the diffusion model (DDPM-DDIM) and DDIM sampling from an image noised with DDIM inversion (DDIM). 

In order to include the four MRI modalities available in BraTS 2021 as inputs to the models, we concatenated them over the channel dimension.


\subsection{Counterfactual Examples}
In Fig.~\ref{fig.plotist} and ~\ref{fig.plot} we display examples of healthy images and anomaly maps obtained with the different approaches. We can observe that f-Ano GAN is not able to generate credible counterfactuals and generally produces images of poor quality and unrealistic appearance. On the other hand, the other approaches are able to create more high-quality results. 

However, in the ones obtained with DenoisingAE, AnoDDPM and AutoDDPM the pathological lesion is still partially visible, while the counterfactuals obtained with CG and CFG seem to present some artifacts, which may not only impact the realism of the counterfactual examples but also the precision of the anomaly maps obtained from them. 
In order to better quantify the capability of these methods to segment pathological areas accurately, we compute the Dice and IoU scores of the anomaly maps they generate.

We also test our approach on healthy samples. Ideally, we would like our generative process to act as the identity function when given a normal image as input. Some examples are shown in Fig.~\ref{fig.healthy}, where we can observe that the changes introduced by our sampling technique are relatively minimal and Dif-fuse preserves the structure and general appearance of the images.

\subsection{Hyperparameters}

In early experiments we observed that, when using the saliency maps to generate the masks needed in Dif-fuse, binarising thems produces better results. Therefore, on the validation set, we explore the optimal thresholding level for the binarisation of the saliency maps and the most appropriate noise amount to employ during the sampling from our diffusion model. 

\begin{figure*}[ht]
\centering
\hspace{1cm}
\begin{subfigure}{0.195\textwidth}
\includegraphics[width=1\textwidth]{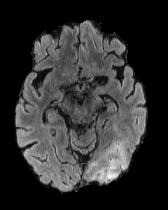}\\
\end{subfigure}\hspace{\fill}
\begin{subfigure}{0.195\textwidth}
\includegraphics[width=1\textwidth]{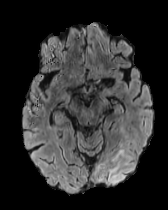}\\
\end{subfigure}\hspace{\fill}
\begin{subfigure}{0.195\textwidth}
\includegraphics[width=1\textwidth]{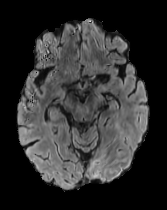}\\
\end{subfigure}\hspace{\fill}
\begin{subfigure}{0.195\textwidth}
\includegraphics[width=1\textwidth]{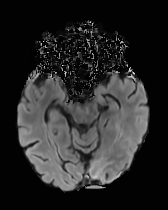}\\
\end{subfigure}\hspace{\fill}

\hspace{1cm}
\begin{subfigure}{0.195\textwidth}
\includegraphics[width=1\textwidth]{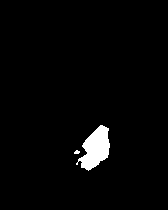}
\caption{Image}
\end{subfigure}\hspace{\fill}
\begin{subfigure}{0.195\textwidth}
\includegraphics[width=1\textwidth]{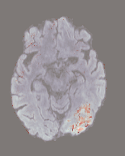}
\caption{K=250}
\end{subfigure}\hspace{\fill}
\begin{subfigure}{0.195\textwidth}
\includegraphics[width=1\textwidth]{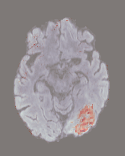}
\caption{K=500}
\end{subfigure}\hspace{\fill}
\begin{subfigure}{0.195\textwidth}
\includegraphics[width=1\textwidth]{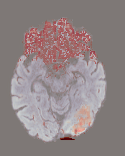}
\caption{K=750}
\end{subfigure}\hspace{\fill}

\caption{Original image with ground truth segmentation mask from BraTS 2021 (a) and healthy images (top row) with corresponding anomaly maps (bottom row), obtained with Dif-fuse with 250 (b), 500 (c), and 750 (d) noising steps. When employing lower amounts of noise, the pathological regions are not fully removed, while when the noise level is too high significant artifacts may be introduced.}
\label{fig.noises}
\end{figure*}

In Fig.~\ref{fig:hyp} we plot the dice scores obtained for different values of these hyperparameters. As we can observe, we obtain the best performance when employing 500 noising steps and selecting the pixels in the $90^{th}$ percentile of the saliency maps. 
In Fig.~\ref{fig.noises} we display counterfactuals obtained with different noise levels. We can observe how smaller values of the noise parameter don't allow the diffusion model to modify the image to an adequate degree, while bigger values introduce artifacts that impact the image quality of the generated image, consequently also hurting the dice score of the corresponding anomaly map.

\subsection{Quantitative Evaluation}

\begin{table*}[t]
\begin{center}
\caption{Dice and IoU scores on BraTS 2021 and WMH test data, KID on IST-3, averaged over three runs (with standard error). Dif-fuse achieves the best anomaly segmentation performance on both BraTS 2021 and WMH. The DDIM inversion followed by DDPM sampling ablation has the best KID on IST3.}\label{tab:dice}

\begin{tabular}{|l|cc|cc|c|}
\hline
\multicolumn{1}{|c|}{Method} & \multicolumn{2}{c|}{BraTS 2021} & \multicolumn{2}{c|}{WMH} & \multicolumn{1}{c|}{IST-3}\\
\multicolumn{1}{|c|}{} & Dice $\uparrow$ & IoU $\uparrow$ & Dice $\uparrow$ & IoU $\uparrow$ & KID $\downarrow$\\
\hline
f-Ano GAN & 0.545 (0.015)   & 0.473 (0.013) & 0.172 (0.022) & 0.103 (0.018) & 0.284 (0.005)\\
Classifier guidance & 0.650 (0.004) & 0.577 (0.003) & 0.468 (0.008) & 0.434 (0.007) & 0.082 (0.002) \\
Classifier-free guidance & 0.631 (0.005) & 0.551 (0.005) & 0.422 (0.009) & 0.354 (0.006) & 0.046 (0.001)\\
AnoDDPM & 0.494 (0.020) & 0.488 (0.017) & 0.151 (0.010) & 0.091 (0.008) & 0.192 (0.022)\\
AutoDDPM & 0.655 (0.007) & 0.584 (0.005) & 0.503 (0.007) & 0.496 (0.005) & 0.073 (0.007)\\
DenoisingAE & \underline{0.681} (0.011) & \underline{0.614} (0.007) & 0.439 (0.015) & 0.370 (0.012) & 0.204 (0.017)\\
Dif-fuse (Ours) & $\mathbf{0.699}$ (0.004) & $\mathbf{0.620}$ (0.004) & $\mathbf{0.569}$ (0.008) & $\mathbf{0.526}$ (0.006) & 0.040 (0.003)\\
\hline
 \multicolumn{6}{|c|}{Ablation experiments} \\
 \hline
ACAT & 0.591 (0.007) & 0.531 (0.005) & \underline{0.530} (0.007) & \underline{0.497} (0.006) & 0.058 (0.002)\\
DDPM & 0.581 (0.003) & 0.501 (0.003) & 0.475 (0.015) & 0.436 (0.012) & \underline{0.039} (0.004)\\
DDIM-DDPM & 0.616 (0.006) & 0.543 (0.007) & 0.498 (0.013) & 0.459 (0.011) & $\mathbf{0.037}$ (0.004)\\
DDIM & 0.498 (0.009) & 0.489 (0.006) & 0.495 (0.009) & 0.490 (0.011) & 0.117 (0.008)\\
DDPM-DDIM & 0.677 (0.004) & 0.605 (0.003) & 0.487 (0.015) & 0.460 (0.013) & 0.085 (0.006)\\
\hline
\end{tabular}

\end{center}
\end{table*}

We evaluate the anomaly maps obtained with the different approaches on BraTS 2021 and WMH. The results are displayed in Table~\ref{tab:dice}. We can observe how our approach obtains the best performance on WMH (with mean Dice and IoU of 0.569 and 0.526 respectively), and BraTS 2021 with 0.699 Dice and 0.620 IoU (with DenoisingAE being second-best on Brats 2021 with Dice and IoU of 0.681 and 0.614 respectively, and ACAT being second-best on WMH with Dice 0.530 and IoU 0.497).

The ablation on the saliency maps obtained from ACAT, that are employed as part of our approach, displays how sampling from the diffusion model as in Dif-fuse is critical to obtain the best results and improve the lesion detection capability of the saliency maps.
Additionally, the ablation on the different combinations of DDPM and DDIM for forward and backward sampling, shows how the combination of both at each sampling step introduced in our approach, together with the masking guidance, are important to achieve the best results.
We have also ablated our method on BraTS 2021 using saliency masks obtained with Grad-CAM~\cite{selvaraju2017grad} and the gradient method~\cite{simonyan2013deep} to guide the sampling from the diffusion model. In particular, with the former approach, we obtained a mean Dice of 0.539 and a mean IoU of 0.512, while with the latter 0.576 and 0.533 respectively. As expected, the results were inferior to the ones obtained with the masks obtained with ACAT (Dice: 0.699, IoU: 0.620) due to the lower quality of these saliency maps, which is consistent with the findings in ACAT.

In Table~\ref{tab:dice} are also displayed the KID scores obtained on IST-3, comparing the generated normal images with real negatives from the dataset.
We selected this metric because it reduces the bias inherent in the Fréchet Inception Distance~\cite{heusel2017gans}, particularly when working with a small number of samples. We compute it using features from the last convolutional layer of the Inception v3 model.
We can observe how the DDIM inversion followed by DDPM sampling ablation has the best KID on IST3 (0.037), followed by DDPM (0.039) and Dif-fuse (0.040). This can be explained with the fact that unconstrained sampling (without masking) as in the ablations can achieve more realistic-looking images. However, it also has the downside of modifying the overall anatomy of the samples, resulting in worse segmentation of the anomaly maps.  

To provide context for our results, it's important to consider the performance of state-of-the-art supervised segmentation methods. On the BraTS2021 test data, the best supervised method\footnote{\url{https://www.synapse.org/Synapse:syn25829067/wiki/611504}.} achieved Dice scores of 0.837, 0.877, and 0.925 for the "enhancing tumor" (ET), "tumor core" (TC), and "whole tumor" (WT) classes, respectively. For WMH data, the top-performing supervised approach achieved a Dice score of 0.81\footnote{\url{https://wmh.isi.uu.nl}.}.
While our method doesn't yet match these supervised results, it demonstrates competitive performance without requiring annotations. This highlights the potential of generative approaches in medical image analysis, especially in scenarios where annotated data is scarce or expensive to obtain.

\subsection{Comparison with Inpainting Methods}

While our proposed method shares similarities with inpainting techniques, there are two key differences. 1) Unlike traditional inpainting, which assumes a predefined mask for the area to be modified, our approach addresses the challenge of identifying the target region automatically, including accounting for the inherent location uncertainty. 2) Inpainting typically involves completing entirely missing sections using only contextual cues. In contrast, our method leverages existing pathological features, which we aim to render as healthy tissue.
These differences necessitate a more nuanced approach that combines elements of inpainting with specialised techniques for medical image analysis and transformation.

As a representative of inpainting approaches, we test Repaint \cite{lugmayr2022repaint} employing the masks obtained with ACAT (as the original method assumes the availability of ground truth masks of the regions that have to be inpainted).
We use 250 timesteps, with 10 times resampling with jumpy size of 10, as recommended in \cite{lugmayr2022repaint}. We obtained Dice score of  0.649 and IoU of 0.575 on BraTS2021, and Dice of 0.532 and IoU of 0.484 on WMH.

It's worth noting that inpainting methods can struggle in our setting as they are not designed to leverage existing information in the masked region or handle uncertainty regarding the area to be inpainted.

\section{Conclusion}
In this work, we propose a method to remove lesions from pathological images through diffusion models, in order to generate credible counterfactuals and produce anomaly maps. To achieve this goal, we employ a two-step approach. First, we utilise ACAT to generate initial saliency maps. These maps provide a first approximation of the areas that require modification. Next, we introduce a novel way to sample from diffusion models. This technique enables us to make targeted modifications to the identified regions while preserving the remaining parts of the image. We fuse both components at each timestep to ensure a smooth transition between the edited and unedited parts and a realistic output. In particular, we inpaint ROIs with DDPM sampling and reconstruct the normal anatomy with DDIMs. By applying some post-processing steps to the difference between the counterfactual example and the original image, we can also obtain the final anomaly map. We observe that our sampling approach not only produces highly realistic counterfactual images but also enhances the initial saliency maps generated by ACAT in the first step.  In particular, we obtain the highest mean Dice and IoU scores of all the methods considered on both BraTS 2021 and WMH, while achieving lower but comparable KID on IST-3 to the unconstrained (without masking) diffusion sampling methods.
Our model demonstrates promising generalisation capabilities across datasets with visually similar pathologies (BraTS2021 and WMH). This cross-dataset performance suggests potential for broader applicability. However, we acknowledge that a full assessment of the generalizability of our approach, particularly to rare or unseen diseases, warrants further exploration. The binary classifier used to compute initial saliency maps is a key component in this regard. To enhance the model's versatility, future work could focus on training this classifier on a more diverse range of pathologies. This would shed light on, and likely improve, the model's ability to identify and process a wider spectrum of anomalies, potentially extending its applicability.
We applied our approach to MRI and CT scans of the brain, but we believe that it can also be employed in many other medical imaging applications where image segmentation is required. We leave further testing for future work.
\appendices

\end{document}